\documentstyle{amsart}
\newtheorem{thm}{Theorem}[subsection]

\newtheorem{defn}[thm]{Definition}
\newtheorem{conj}[thm]{Conjecture}

\newenvironment{pr}{{\em Proof}\newline }{\epr}

\numberwithin{equation}{section}

\begin{document}
%
%
\newcommand{\bC}{{\Bbb C}}
\newcommand{\bN}{{\Bbb N}}
\newcommand{\bQ}{{\Bbb Q}}
\newcommand{\bR}{{\Bbb R}}
\newcommand{\bZ}{{\Bbb Z}}
%
%
\newcommand{\cA}{{\cal A}}
\newcommand{\cB}{{\cal B}}
\newcommand{\cC}{{\cal C}}
\newcommand{\cD}{{\cal D}}
\newcommand{\cE}{{\cal E}}
\newcommand{\cF}{{\cal F}}
\newcommand{\cG}{{\cal G}}
\newcommand{\cH}{{\cal H}}
\newcommand{\cI}{{\cal I}}
\newcommand{\cJ}{{\cal J}}
\newcommand{\cL}{{\cal L}}
\newcommand{\cM}{{\cal M}}
\newcommand{\cN}{{\cal N}}
\newcommand{\cO}{{\cal O}}
\newcommand{\cP}{{\cal P}}
\newcommand{\cR}{{\cal R}}
\newcommand{\cS}{{\cal S}}
\newcommand{\cT}{{\cal T}}
\newcommand{\cU}{{\cal U}}
\newcommand{\cX}{{\cal X}}
%
%
\newcommand{\gog}{{\frak{g}}}
\newcommand{\goh}{{\frak{h}}}
\newcommand{\gon}{{\frak{n}}}
%
%
\newcommand{\ub}{{\underline b}}
\newcommand{\uc}{{\underline c}}
\newcommand{\ud}{{\underline d}}
\newcommand{\ue}{{\underline e}}
\newcommand{\uG}{{\underline G}}
\newcommand{\ui}{{\underline i}}
\newcommand{\uI}{{\underline I}}
\newcommand{\uJ}{{\underline J}}
\newcommand{\uL}{{\underline L}}
\newcommand{\uU}{{\underline U}}
\newcommand{\uQ}{{\underline Q}}
%
%
\newcommand{\ta}{{\tilde a}}
\newcommand{\tc}{{\tilde c}}
\newcommand{\te}{{\tilde e}}
\newcommand{\tj}{{\tilde j}}
%
%
\newcommand{\trr}{\triangleright}
\newcommand{\epr}{\begin{flushright} $\Box$ \end{flushright}}
\begin{center}
{\Large \bf Concerning a natural compatibility condition between\\
the action and the renormalized operator product}\\
Olivier de Mirleau\\
\hspace{2mm}mirleau@@wins.uva.nl
\footnote{
Department of Mathematics/
University of Amsterdam/
Plantage Muidergracht 24/
1018 TV Amsterdam/
The Netherlands
}\\
\end{center}
\tableofcontents
\section{Abstract}
In this article we note that in a number of situations the operator
product and the classical action satisfy a natural compatibility condition.
We consider the interest of this condition to be twofold:
First, the naturality (functoriality) of the compatibility condition
suggests that it be used for geometrical applications of renormalized
functional integration.
Second, the compatibility can be used as the definition of a category;
consideration of this category as the central object of study in
quantum field theory seems to have quite some advantages over previously
introduced theories of the type ``S-matrix theory'',
``Vertex operator algebras'', since this seems to be the only
category in which both the action and the expectation values enter,
the two being linked roughly speaking by a combination of the Frobenius
property and the renormalized Schwinger-Dyson equation.
\newpage
\section{Introduction.}
Functional integrals are a very compact way of expressing expectation values
relevant for the description of physical phenomena in terms of an action,
but they have a disadvantage, being that
they do not contain enough information: The action has to be
supplemented with renormalization conditions before expectation values
can be determined.
This article aims at providing a setting in which this kind of information
can be handled more systematically.

This leads us to an interesting
axiomatization of renormalized quantum field theory in general, not
restricting to a certain action-independent aspect of it like
the gluing property, or the presence of a certain symmetry.
This axiomatization seems adequate as a setting for quantum field
theory because it requires a link between the action and the expectation
values, unlike for example S-matrix theory \cite[chapter 2-3]{chew},
the theory of $n$-point functions satisfying Wightman's axioms
\cite[chapter 3]{streater},
Segal's gluing functors \cite{segal}\cite{atiyah}, or the theory
of vertex operator algebras \cite{frenkel}.

The article is organized as follows:
In section \ref{renosec} we define renormalization and illustrate it with
a number of examples. In section \ref{normosec} we first remind the link
between normal ordering and renormalization, and then prove that a natural
compatibility condition with the action is satisfied by normal ordering
for actions of which the first derivatives constitute a coordinate system.
In section \ref{restartsec} we define the category of renormalized volume
manifolds, which axiomatize our compatibility condition,
and illustrate them with some easy examples. The appendix
contains a review of functional integration, partially overlapping with
contents of \cite{mirleau}.

\newpage
\section{Renormalization.}
\label{renosec}
\subsection{The general idea.}
The general idea of renormalization is that if a function $G(\beta,\Lambda)$
of several variables diverges as one of the variables approaches a certain
limit, say $\Lambda\rightarrow\infty$,
then this divergence can sometimes be ``compensated'' or ``removed'' by
simultaneously adapting the other variables, say
$\beta\rightarrow-\infty$. I.e. $\beta$ is replaced by a function
of $\Lambda$ such that $\lim_{\Lambda\rightarrow\infty}
\;G(\beta(\Lambda),\Lambda)$ is non-empty.

In general of course this limit will depend on the choice of function
$\beta(.)$.
One may however ask exactly \underline{what set} of
values can be obtained in this way by varying the function $\beta(.)$, i.e.
we may try to specify the set:
$$\cup_{\beta(.)} \lim_{\Lambda\rightarrow\infty}
\;G(\beta(\Lambda),\Lambda).$$

This expression can be simplified by eliminating reference to the functions
$\beta(.)$: Indeed, if $\cB$ is the set of values for $\beta$,
then such limit points are in the closure
$\overline{G(\cB,(n,\infty))}$, for any $n$.
Conversely, under general assumptions
\footnote{Let the space $\cE$ of \ref{rendefn} be first countable, $p\in
\lim G$, then there are $\beta_n,\lambda_n$ such that
$G(\beta_n,\lambda_n)\rightarrow p$ and $\lambda_n\rightarrow\infty$.
Restricting to an increasing subsequence of the $\lambda_n$'s, we may
extend it to an invertible $x\mapsto \lambda_x$ with
$\lim_{x\rightarrow\infty} G(\beta_{\lceil x \rceil},\lambda_x)=p$, and
since $\lambda$ is invertible this implies that $p$ is a limit point of the
first form.}
one may prove that this is equivalent
to the previous expression.
This leads to the following definition:
\subsection{The definition of renormalization.}
We propose the following as a general definition of renormalization;
We will illustrate it with examples below:
\begin{defn}
\label{rendefn}
Let three topological spaces be given:
\begin{enumerate}
\item $\cB$, the space of Bare parameters.
\item $\cC$, the space of Cutoffs.
\item $\cE$, the space of Expectation values, or of possible Experimental
values.
\end{enumerate}
Let $\Lambda_\infty\in \cC$ be a point in the cutoff space,
and suppose given a function depending on bare parameters and cutoffs,
which is not defined at $\Lambda_\infty$:
$$G:\cB\times(\cC-\{\Lambda_\infty\})\rightarrow \cE.$$
Then:
\begin{enumerate}
\item We define the limit $\Lambda\rightarrow \Lambda_\infty$ of the
sets $G(\cB,\Lambda)$ as follows:
$$\lim G:=\lim_{\Lambda\rightarrow\Lambda_\infty}(G(\cB,\Lambda)):=
\cap_{\cU\;neighborhood\;of\;\Lambda_\infty}
\;\;\overline{G(\cB,\cU-\{\Lambda_\infty\})}.$$
\item We say that $G$ is renormalizable iff $\lim G$ is non-empty.
\footnote{For applications to the description of physical phenomena
it is common to strengthen this definition by
requiring that the limit be finite-dimensional; The idea is that all
experimental data together is roughly represented by a \underline{point}
$e\in \cE$, whereas
$G$ determines a limiting \underline{set} in $\cE$:
A possible description of $e$ would be to just give all its
coordinates in $\cE$, but since $\cE$ is very big this is not a practical way.
It turns out that we can describe $e$ in a very satisfactory way by first
narrowing it down to be in a finite dimensional limit set of some
relatively simple $G$, and then fixing $e$ by giving a finite number of
coordinates of $e$ (the renormalization conditions) along with $G$.
Infinite dimensional limiting sets are not very useful for this purpose
since one would have to specify an infinity of experimental results
to pinpoint $e$.}
\item A set of locally defined functions
$\rho_{i\in I}:\cE\rightarrow \bR$ such that their
restriction to $\lim G$ is a local bijection $\lim G\rightarrow \bR^I$
will be called a set of renormalized parameters.
\end{enumerate}
\end{defn}
\subsection{Two examples.}
We will now consider two examples of renormalization. In the first
example, $\cE$ will be two-dimensional and one may draw a
picture of what is happening.
The second example will be less suitable for pictures,
but it is an example of renormalizing divergent integrals,
which is what we will concentrate on later.
\subsubsection{Low dimensional example of renormalization.}
As an illustration, consider the following case:
$$\cB:=\bR;\;\;\cC:=(0,\infty];\;\;\cE:=\bR^2;\;\;
G(\beta,\Lambda\neq\infty):=(\beta+\Lambda,1-\Lambda^{-1}).$$
Note that $G(\beta,\infty)$ is not defined.
But if we consider all $\beta$'s at the same time,
then the limit $\Lambda\rightarrow\infty$ can be defined:
$$G(\bR,\Lambda)=(\bR+\Lambda,1-\Lambda^{-1})=(\bR,1-\Lambda^{-1})
\rightarrow (\bR,1).$$
%
The sense in which we consider this
set to be the limiting set is exactly stated in our general definition:
Indeed, we have
$$\cap_{n\in \bN}\overline{G(\cB, (n,\infty))}
=\cap_{n\in \bN}\overline{\bR\times (1-n^{-1},1)}
=\cap_{n\in \bN} \bR\times [1-n^{-1},1]
=(\bR,1).$$

Thus, $G$ above is renormalizable with limiting set
$(\bR,1)$; the first coordinate is a renormalized parameter for $G$.

We have seen that the divergence
$\lim_{\Lambda\rightarrow\infty} G(\beta,\Lambda)$ has been removed
by taking all $\beta$'s together. Another common way of saying this
is that the divergence has been ``absorbed'' into the bare
parameters in the step $\bR+\Lambda=\bR$.

Note the relation between the renormalized parameters, and the absorption
of the divergence: Indeed, setting $\beta_\rho(\Lambda):=\rho-\Lambda$,
we have
$$G(\beta_\rho(\Lambda),\Lambda)=
(\beta_\rho(\Lambda)+\Lambda,1-\Lambda^{-1})=(\rho,1-\Lambda^{-1})
\rightarrow (\rho,1).$$

I.e. we can also think of the renormalized parameters as parametrizing
a set of functions $\beta(.)$ such that the different possible limiting
values of $G$ are exactly obtained by substituting these functions $\beta(.)$
for the variable $\beta$.
\subsubsection{An example of renormalized divergent integration.}
For the next example, set
$$\cB:=\bR^{>0}\times \bR^{>0};\;\cC:=(0,\infty];\;\cE:=\bR^\bN=Map(\bN,\bR),$$
$$G(\beta_1,\beta_2;\Lambda):=
\{n\mapsto \int_0^\Lambda (\beta_1 x)^n (\beta_2 x) dx\}.$$
In other words:
$$G(\beta,\Lambda)=\{n\mapsto \beta_1^n \beta_2 \Lambda^{n+2}/(n+2)\}
=(\beta_2 \Lambda^2/2,\beta_1 \beta_2\Lambda^3/3,\beta_1^2\beta_2 \Lambda^4/4,
...)$$
The set $G(\cB,\Lambda)$
is a two-dimensional subset of $\bR^\bN$, which
we may try to parametrize with the first two coordinates, which we call
$\rho_1$ and $\rho_2$. Thus, on the set $G(\cB,\Lambda)$
we have the following relations:
$$\rho_1=\beta_2 \Lambda^2/2;\;\;\rho_2=\beta_1 \beta_2\Lambda^3/3,$$
so that the $\rho$'s are positive, since the $\beta$'s are. These
relations can be inverted, giving
$$\beta_1={3\rho_2\over 2\rho_1\Lambda};\;\;\beta_2={2\rho_1\over \Lambda^2},$$
which we can in turn substitute in $G$.
Thus, $(\rho_1,\rho_2)\in\cR:=\bR^{>0}\times\bR^{>0}$, are
renormalized parameters
since with $F:\cR\rightarrow \cE=\bR^\bN$:
$$F(\rho_1,\rho_2):=\{n\mapsto (2\rho_1)^{1-n} (3\rho_2)^n /(n+2)\}
=(\rho_1,\rho_2,{9 \rho_2^2\over 8\rho_1}, ...),$$
we see that for all $\Lambda$: $G(\cB,\Lambda)=F(\cR)$, and in particular
$F(\cR)=\lim_{\Lambda\rightarrow \infty} G(\cB,\Lambda)$.
Other renormalized parameters are given
by $(\rho_2,\rho_3)$, corresponding to the parametrization
$$F(\rho_2,\rho_3):=\{n\mapsto (3\rho_2)^{2-n} (4\rho_3)^{n-1}/(n+2)\}
=
({9\rho_2^2\over8\rho_3},\rho_2,\rho_3,
{16\rho_3^2\over 15\rho_2},...).$$
\subsection{Non-generic examples.}
In the above example we have seen that $\lim G$ was
homeomorphic to the set of bare parameters. This is not true in general:
We will now have a look at some examples where the topology or even the
dimension of $\cB:=\bR$ and $\cR:=\lim G$ differ.
\footnote{In the context of renormalized functional integration,
the case where
$\dim(\cR)<\dim(\cB)$ and where ``what is left'' corresponds to Gaussian
expectation values, is referred to as triviality, see for example
\cite[section 2]{luescher}.}

First, consider
$$G(\beta, \Lambda):=(\cos(\beta),\sin(\beta)/\Lambda).$$
The limiting set equals $[-1,1]\times \{0\}$, so that $\cR\cong [-1,1]$,
which has topology different from $\cB=\bR$. If we replace
$\cos(\beta)$ by $\cos(\beta)/\Lambda$, then $\cR$ is just a point.

Finally, the next example will be a case where the dimension of $\cR$ is bigger
than that of $\cB$: Indeed, set
$$G(\beta,\Lambda):=e^{-\beta-\Lambda}(\cos(\beta),\sin(\beta)).$$
For fixed $\Lambda$ this describes a spiral in $\bR^2$. We claim that
for all $n$: $G(\bR,(n,\infty))=\bR^2-\{0\}$; Indeed, any point in
$\bR^2-\{0\}$ can be written as
$$e^{-d}(\cos(\phi),\sin(\phi))=G(\phi-2\pi k,d-\phi+2\pi k)
\in G(\bR,(n,\infty)),$$
if we choose $k$ big enough.
Thus, the limiting set is equal to $\bR^2$, so that $\cR$ has dimension
higher than $\cB$.
\newpage
\section{Two properties of normal ordering.}
\label{normosec}
\subsection{Reminder on normal ordering.}
\subsubsection{Definition.}
Let $S$ be a function on $\bR^n$ such that its first derivatives
$S_i:=\partial_i S$
form a coordinate system. Normal ordering acts on polynomial functions of these
coordinates and gives back functions on $\bR^n$; It is defined inductively
as follows \cite{mirleau}:
$N(1):=1$, and
$$N(S_{i_0}..S_{i_n}):=S_{i_0} N(S_{i_1}..S_{i_n})-\frac{\partial}{
\partial x^{i_0}}
N(S_{i_1}..S_{i_n}).$$
\subsubsection{Relation with Schwinger-Dyson equation.}
Normal ordering together with the Schwinger-Dyson equation
$\langle (\partial_i S)f \rangle=\langle \partial_i f \rangle$
(see appendix) implies
$$\langle N((\partial_i S)f)g\rangle=\langle N(f)\partial_i g\rangle.$$
If $N$ is invertible, then
$\{I$ satisfies the Schwinger-Dyson equation $\Leftrightarrow
I(f)=ZN^{-1}(f)I(1)\}$,
where $Z$ denotes the projection of polynomial functions of the $S_i$'s
on their constant part, e.g. $Z(3+aS_1+b S_1 S_5)=3$.
\subsubsection{Relation with original definition.}
Houriet and Kind's original definition \cite[Formula 12]{houriet}
using the creation-annihilation
decomposition of operator fields $\phi=\phi^-+\phi^+$ amounts to
$$(:1:):=1;\;(:f\phi:):=\phi^-(:f:)+(:f:)\phi^+.$$
It satisfies properties
analogous to those of $N$ if we replace the operation $\partial_i$
with the commutator $[\phi^+,.]$:
\begin{enumerate}
\item $(:\phi_1..\phi_n:)=(:\phi_{\sigma(1)}..\phi_{\sigma(n)}:)$.
\label{normprop1}
\item $(:\phi f:)=\phi (:f:)-[\phi^+, (:f:)]$.
\label{normprop2}
\item $\langle 0|(:\phi f:) g|0\rangle
=\langle 0| (:f:) [\phi^+, g] |0\rangle$,
\end{enumerate}
(see \cite[Appendix A.4]{mirleau} for more details). This motivates the use
of the name normal ordering for $N$.
\subsubsection{Relation with substraction of singularities.}
In infinite dimensions, the partial derivatives $\partial_i$ get replaced
by functional derivatives, and solutions of the corresponding
Schwinger-Dyson equation exhibit so-called UV-divergences, i.e. short
distance singularites as in $\langle \phi(x)\phi(y)\rangle=
K|x-y|^{2-D}$. This makes it impossible to take expectation values of say
$\phi^2(x)$. However, normal ordering turns out to be a way to
substract one singularity from the other such as to
leave something finite:
Let us restrict the attention to Gaussian normal ordering; then in infinite
dimensions we have $N(\phi(x)\phi(y))=\phi(x)\phi(y)-
\langle \phi(x)\phi(y)\rangle$, and we see that $N(\phi^2(x))$ is only
defined if the two-point function singularity is regulated, say by some cutoff
parameter $\Lambda$; $N$ will in general depend on $\Lambda$ and the
particular regulator; The inductive definition now amounts to:
$$N_\Lambda(1):=1;\;\;N_\Lambda(\phi^{n+1}(x)):=
\phi(x)N_\Lambda(\phi^n(x))-n\langle \phi(x)\phi(x)\rangle_\Lambda
N_\Lambda(\phi^{n-1}(x)).$$
The central statement relating $N$ to
the substraction of short distance singularities is that
$\langle \prod_{j=1}^n N(e^{p_j\phi(x_j)})\rangle
=\prod_{i<j} e^{p_i p_j \langle \phi(x_i)\phi(x_j)\rangle}$, which is finite
for $x_i\neq x_j$. By taking derivatives w.r.t. $p$ we see that expectation
values of products of expressions like $N(\phi^n(x))$ are finite, i.e. we
see that normal ordering exactly substracts singularities coming from
$\phi^n(x)$. Furthermore, note that the final result of this substraction
is independent of the particular regulator used as the cutoff is removed.
\subsection{Property one: Gaussian normal ordering by renormalization.}
In this section we will further illustrate the renormalization procedure
by showing how Gaussian normal ordering is a case of renormalization.
As we have  presented it above,
the limit $\Lambda\rightarrow \infty$ cannot be renormalized since
we have not specified bare parameters in which to absorb the divergences.
This is what is done below:
\begin{defn} Define spaces $\cB$ and $\cE$ as follows:
\begin{enumerate}
\item The set of bare parameters $\cB$ will be
the space of infinite triangular matrices with real entries:
$\beta=(\beta_{nm})_{n=1..\infty, m=0..n-1}$.
\item To every bare matrix $\beta$ is associated a linear operator $R$
on polynomials of one variable $x$, as follows:
$$R_\beta(x^n):=x^n+\sum_{i=0}^{n-1} \beta_{ni} x^i.$$
\item Let $\cE$ be the following space of arrays of functions:
$$\cE:=\{ E=(E^{n_1,..,n_k})_{n_i=1..\infty, k=0 ..\infty}|
E^{n_1,..,n_k}:(\bR^D)^k\rightarrow \bR
\}.$$
\item With $\cC:=(0,\infty]$, let $G:\cB\times (\cC-\infty)
\rightarrow \cE$ be the map which to bare matrix $\beta$ and cutoff
$\Lambda$ associates the array of functions $G^{n_1,..,n_k}$ given by
$$G^{n_1,..,n_k}(x_1,...,x_k):=
\langle R_\beta(\phi^{n_1}(x_1)) ... R_\beta(\phi^{n_k}(x_k) \rangle_\Lambda.$$
\item Finally, for $n=1...\infty$ and $m=0...n$, define functions
$\rho_{nm}:\cE\rightarrow \bR$
by
$$\rho_{nm}(E):=E^{n,1,..,1}(0,
\stackrel{m\;times}{\overbrace{0,...,0}}),$$
i.e. we have
$$\rho_{nm}(G(\beta,\Lambda))=
\langle R_\beta(\phi^n(0))
\stackrel{m\;times}{\overbrace{R(\phi(0))...R(\phi(0))}}\rangle.$$
\end{enumerate}
\end{defn}
The precise sense in which normal ordering is related to renormalization is
the following:
\begin{thm} The above $G$ is renormalizable; More specifically,
there is a limit
point satisfying the renormalization condition
$$\rho_{(n+1),m}=\sum_{i=1}^m \rho_{11}\rho_{n,(m-1)}.$$
It is reached by defining the $\Lambda$-dependence of the bare parameters
to be given by normal ordering: $R_{\beta(\Lambda)}:=N_\Lambda$.
\end{thm}
\begin{pr}
If $R_{\beta(\Lambda)}:=N_\Lambda$, then the renormalization condition
is satisfied for all $\Lambda$ since normal ordering satisfies
$$
\langle N(\phi^{n+1}(0))
\stackrel{m\;times}{\overbrace{N(\phi(0))...N(\phi(0))}}\rangle$$
$$=
\sum_{i=1}^m
\langle N(\phi(0)) N(\phi(0)) \rangle
\langle N(\phi^n(0))
\stackrel{m-1\;times}{\overbrace{N(\phi(0))...N(\phi(0))}}\rangle.$$
$G$ is renormalizable because the cutoff-limit of expectation values
of normal ordered integrands exists.
\end{pr}
\subsection{Property two: Natural compatibility with the action.}
\subsubsection{Normal ordering of vectorfields.}
The defining formula
$N((\partial_i S)f)=(\partial_i S) N(f)-\partial_i N(f)$ has a number of
unsatisfactory features:
\begin{enumerate}
\item The transition between $x^i$'s and $\partial_i S$'s may not be
invertible; It might therefore be that some functions cannot be written as
a function of $\partial_i S$'s.
\item The
definition needs the choice of vectorfields $\partial_i$; On more
general manifolds the choice of a subalgebra of vectorfields is needed.
One would like to have a more natural notion of substracting short
distance singularities, not involving an a priori choice of vectorfields.
\end{enumerate}
These considerations, and the
hope of simplifying a number of formulae have led us to introduce
normal ordering of vectorfields
by $N(X^i\partial_i):=N(X^i)\partial_i$:
The combination of $N$, $S$ and $\nabla_S$
\footnote{With $\mu_S:=e^{-S}dx^1 ... dx^n$, $\nabla_S$ is the divergence of
$\mu_S$: $\nabla_S(X)\mu_S:=L_X\mu_S$.}
then satisfies the following attractive compatibility condition:
(We will prove it in a moment):
$$\nabla_S(N(X))=-N(X(S)).$$
Indeed, this equation has the following features:
\begin{enumerate}
\item The vectorfields $X$ are not restricted to be of the form $\partial_i$.
In other words, the \underline{naturality} of this expression is higher
than the seperate defining formulae of $N(f)$ and $N(X)$.
This is the naturality that we are referring to in the title of this paper.
\item If we know $N$ and $S$ then we know $\nabla_S$, i.e. it makes the
relation between divergence $\nabla$ and regulator $N$ explicit once the
action is known. Therefore, if we require this compatibility to hold
\underline{outside} our restricted class of actions, then in general
for nonlinear targetspaces, the problem
of regularizing composite operators and that of defining $\nabla_S$ are
the same.
\item The Schwinger-Dyson equation can now be stated without $\nabla_S$,
instead using only $N$: $[N(X(S))f]=[N(X)(f)].$
\end{enumerate}

With some hindsight, we can reformulate the above as follows:
There are two occasions on which the sole datum of an action $S$ is
not enough to specify expectation values:
\begin{enumerate}
\item $S$ does not determine what we mean by
$\langle R(\phi^2(x))\phi(y)\phi(z)\rangle$, since $R$ as such is not
determined by the action alone. It is only fixed after
\underline{extra conditions} are given, one of these conditions
for example leading to $R=N$, normal ordering.
\item If target space is nonlinear, then $\langle.\rangle$ is
undetermined by an action, since the very statement of the Schwinger-Dyson
equation needs a \underline{divergence} $\nabla$, not an action $S$.
(See appendix \ref{volmansec}).
\end{enumerate}
We now see explicitly that these two matters are related since the
compatibility condition relates $N$ and $\nabla_S$.
>From a more general standpoint this can be understood as follows:
In the case of nonlinear targetspaces,
there is no distinction between expressions that are linear and those are not.
But non-linear arguments like $\phi^2(x)$ are exactly those that need
regularizing.
Thus, the nonlinear targetspaces
are expected to need ultraviolet regulators right from the start, so that
at fixed action $S$ one can understand that some information
contained in $\nabla_0$ is reflected 
in information contained in the choice of
ultraviolet regulators.

Let us now prove the compatibility relation:
(It can of course by definition only be proved for
the restricted class of actions where $\partial_i S$'s are coordinates,
but the class seems flexible enough to want to extend the compatibility
to all actions).
\begin{thm} Let $\nabla_0(X)=\partial_i X^i$, $\nabla_S(X)=\nabla_0(X)-X(S)$,
and let $N$ map vectorfields to vectorfields and functions to functions such
that $N(1)=1$. Assume the first derivatives $\partial_i S$ generate the
algebra of functions $f$ under consideration.
Then the following three points are equivalent:
\begin{enumerate}
\item
$$\nabla_S(N(X))=-N(X(S)),$$
$$N(X^i \partial_i)=N(X^i)\partial_i.$$
\item
$$N((\partial_i S)f)=(\partial_i S) N(f)-\partial_i N(f),$$
$$N(\partial_i)=\partial_i;\;\;
N((\partial_i S)X)=(\partial_i S) N(X)-[\partial_i, N(X)].$$
\item
$$N((\partial_i S)f)=(\partial_i S) N(f)-\partial_i N(f),$$
$$N(X^i \partial_i)=N(X^i)\partial_i.$$
\end{enumerate}
\end{thm}
\begin{pr}
\begin{enumerate}
\item $(2\Rightarrow 3)$. Induction on $|X^i|$; Assume true for $X^i$, we
will prove it for $(\partial_j S)X^i$:
$$N((\partial_j S) X^i \partial_i)=
(\partial_j S) N(X^i \partial_i) - [\partial_j, N(X^i\partial_i)]$$
$$=(\partial_j S) N(X^i)\partial_i -[\partial_j, N(X^i)\partial_i]$$
$$=\{(\partial_j S) N(X^i) -\partial_j N(X^i)\}\partial_i
=N((\partial_j S) X^i)\partial_i.$$
\item $(3\Rightarrow 2)$:
$$N((\partial_i S)X)=N((\partial_i S) X^j \partial_j)$$
$$=N((\partial_i S) X^j)\partial_j
=(\partial_i S) N(X^j) \partial_j - (\partial_i N(X^j))\partial_j$$
$$=(\partial_i S)N(X)-[\partial_i, N(X^j)\partial_j]
=(\partial_i S)N(X)-[\partial_i,N(X)].$$
\item $(2+3\Rightarrow 1)$: Induction on $|X|$, first $|X|=0$:
$$\nabla_S(N(\partial_i))=\nabla_S(\partial_i)
=-\partial_i S=-N(\partial_i S).$$
Assume true for $X$, we will prove it for $(\partial_j S) X$:
$$\nabla_S(N((\partial_j S) X)) + N((\partial_j S) X(S))$$
$$=
\nabla_S((\partial_j S) N(X) - [\partial_j, N(X)])
+(\partial_j S) N(X(S)) - \partial_j N(X(S))$$
$$=
(\partial_j S) \nabla_S(N(X))_1
-\partial_j \nabla_S(N(X))_2
+(\partial_j S)N(X(S))_1
- \partial_j N(X(S))_2$$
$$+N(X)(\partial_j S)_3 + N(X)\nabla_S(\partial_j)_3=0.
\;\;\;\;\;\;\;\;\;\;\;\;\;\;\;\;\;
\;\;\;\;\;\;\;\;\;\;\;\;\;\;\;\;\;$$
\item $(1\Rightarrow 3)$. Induction on $|f|$. $f=1$:
$$N(\partial_i S)= -\nabla_S(N(\partial_i))
=-\nabla_S(\partial_i)=\partial_i S.$$
Suppose the relation proved for $f$, we will prove it for $(\partial_j S)f$:
$$N((\partial_j S) f (\partial_i S))=
-\nabla_S(N((\partial_j S) f \partial_i))
=-\nabla_S(N((\partial_j S) f ) \partial_i)$$
$$=-N((\partial_j S)f) \nabla_S (\partial_i )
-\partial_i N((\partial_j S) f)$$
$$=(\partial_i S) N((\partial_j S) f)
-\partial_i N((\partial_j S)f).$$
\end{enumerate}
\end{pr}
These considerations lead us to the following definition of compatibility
between regulators and the action $S$, whatever type $S$ may be of:
\begin{defn}
Let $A$ be a symmetric associative algebra with unit, $L:=Der(A)$.
Let $S\in A$. Let $N:A\rightarrow A$, and $N:L\rightarrow L$ be invertible.
We say that $(S,N)$ is compatible
iff $\nabla$ defined by
$\nabla(N(X))=-N(X(S))$ is a divergence operator, i.e. satisfies
\begin{enumerate}
\item $\nabla([X,Y])=X(\nabla(Y))-Y(\nabla(X))$.
\item $\nabla(fX)=X(f)+f\nabla(X).$
\end{enumerate}
\end{defn}
The associated Schwinger-Dyson equation reads: $[N(X(S))f]=[N(X)(f)],$
the idea being that this is
the nonlinear generalization \underline{and} the ultraviolet
regularization of $[(\partial_i S) f]=[\partial_i(f)]$. In the rest of this
section $N$ will be assumed to satisfy compatibility with $S$.

Note that the compatibility condition as such does not determine
$N$,
since we dropped the extra requirement
$N(X)=N(X^i)\partial_i$ that fixes $N$. The idea is that the condition
$\nabla_S(N(X))=-N(X(S))$ is
very general, whereas an extra condition like
$N(X)=N(X^i)\partial_i$ is special for the situation under consideration.
I.e. the choice of vectorfields $\partial_i$ is fine when working on linear
spaces, but
on nonlinear spaces,
other types of renormalization conditions will have to be imposed if structure
invariance is to be preserved.
\footnote{For example if targetspace is a Riemannian manifold,
a probable way would be to make $N$ act diagonally on the eigenfunctions of the
Laplacian, by analogy with $N(e^{px})=e^{p^2/2}e^{px}$.}

\subsubsection{Renormalized algebraic operations and the renormalized
Schwinger-Dyson equation.}
Finally let us push the abstraction a little further,
in order to remove one last bad property of normal ordering, which is
that it is only defined at finite cutoff: $N_\Lambda(\phi^2(x))$ diverges
in the cutoff limit. This property can be circumvented, by restricting the
attention to expressions of the type $N^{-1}(N(A(x))N(B(y)))$ at $x\neq y$,
since e.g. in the Gaussian case
$$N^{-1}(N(e^{\phi(x)})N(e^{\phi(y)}))=e^{\langle\phi(x)\phi(y)\rangle}
e^{\phi(x)+\phi(y)},$$
which is well defined.
Thus, we are led to define renormalized algebraic operations
$$f._r g:=N^{-1}(N(f)N(g)),\;f._r X:=N^{-1}(N(f)N(X)),$$
and analogously for $X._r f$ and $[X,_r Y]$, and renormalized expectation
values
$$\langle \cO_1,...,\cO_n\rangle_r:=
\langle N(\cO_1)...N(\cO_n)\rangle,$$
and forgetting completely about $N$, by just working with these
new structures only. Note that the new structure will also satisfy
associativitly, Jacobi, etc., since it is the pullback of such operations
by an invertible map. Since we have
$\langle \cO_1,\cO_2,\cO_3\rangle
=\langle \cO_1._r\cO_2,\cO_3\rangle$, we will refer to $f._r g$ as the
operator product \cite[Section II]{wilson} of $f$ and $g$.
Furthermore:
\begin{thm}
The compatibility condition of $N$ with $S$ implies:
$$X._r (YS)-Y._r (XS)=[X,_r Y] S,$$
$$(f._r X)S=f._r (XS)-X._r f.$$
The Schwinger-Dyson equation can be written as:
$$\langle XS,\cO_1,...,\cO_n\rangle_r=
\sum_{i=1}^n  \langle \cO_1,.., X._r \cO_i,..,\cO_n\rangle_r.$$
\end{thm}
\begin{pr}
\begin{enumerate}
\item
$$N(X._r(YS)-Y._r (XS))=N(X)N(YS)-N(Y)N(XS)$$
$$=-N(X)\nabla(N(Y))+N(Y)\nabla(N(X))
=-\nabla([N(X),N(Y)])$$
$$=-\nabla(N([X,_r Y]))=N([X,_r Y]S).$$
\item
$$N((f._r X)S)=-\nabla(N(f._r X))=-\nabla(N(f)N(X))$$
$$=-N(f)\nabla(N(X))-N(X)(N(f))$$
$$=N(f)N(XS)-N(X._r f)=N(f._r(XS) - X._r f).$$
\item Finally, for the Schwinger-Dyson equation:
$$\langle XS,\cO_1\rangle_r=\langle N(XS)N(\cO_1)\rangle
=\langle N(X)N(\cO_1)\rangle$$
$$=\langle N(X._r\cO_1)\rangle=\langle X._r \cO_1\rangle_r.$$
\end{enumerate}
\end{pr}

This motivates us to study the category introduced in section \ref{restartsec}.
\newpage
\section{Compatibility of operator product and action as a starting point.}
\label{restartsec}
\subsection{Renormalized volume manifolds.}
(For a review of volume manifolds, see
appendix \ref{volmansec}).
\begin{defn}
Let $A$ be a symmetric associative algebra with unit, $L:=Der(A)$,
with corresponding operations
$fg$, $fX$, $Xf$ and $[X,Y]$.
By a renormalized structure on $A$ we mean the datum of $S\in A$,
together with extra multiplications
$f._r g$, $f._r X$, $X._r f$ and $[X,_r Y]$,
such that
\begin{enumerate}
\item The $._r$-operations induce $L=Der(A,._r)$, i.e. the derivations of the
multiplication $(f,g)\mapsto f._rg$ are exactly given by the operations
$f\mapsto X._r f$ as $X\in L$.
\item Both structures have the same unit: $1f=f=1._r f$.
\item The two algebraic structures and the action $S$ are compatible in the
sense that:
$$X._r (YS)-Y._r (XS)=[X,_r Y] S,$$
$$(f._r X)S=f._r (XS)-X._r f.$$
\end{enumerate}
$A$ together with the renormalized structure will be called a renormalized
volume manifold.
For symmetric linear maps $\langle,,,\rangle_r: A^{\otimes n}
\rightarrow \bR$ we will be interested in the following properties:
\begin{enumerate}
\item $\langle .,. \rangle_r:A^{\otimes 2}\rightarrow \bR$ is a positive
nondegenerate form. (Positivity).
\item $\langle \cO_1,\cO_2,...,\cO_n\rangle_r=
\langle \cO_1._r \cO_2,..,\cO_n\rangle_r,$ (Frobenius).
\footnote{A Frobenius algebra is a symmetric associative algebra with metric
such that $(a,bc)=(ab,c)$.}
\item$\langle XS,\cO_1,...,\cO_n\rangle_r=
\sum_{i=1}^n  \langle \cO_1,.., X._r \cO_i,..,\cO_n\rangle_r.$
(Schwinger-Dyson).
\end{enumerate}
\end{defn}
By substituting $S/\hbar$ for $S$ and taking the limit $\hbar\rightarrow 0$,
we see that at $\hbar=0$, we may take both algebraic structures to be
equal. In what follows, we will take the usual multiplication to have
priority over the renormalized one, i.e. $f._r gh$ means $f._r (gh)$.
\subsection{Example 1: Renormalization conditions for Gaussian integrals.}
The following theorem says that in the Gaussian case the renormalized product
subjected to the renormalization condition $f\partial_i._r g=
f._r\partial_i g$ equals $e^{px}._r e^{qx}=e^{\hbar pq} e^{(p+q)x}$:

\begin{thm} Let $A$ be the polynomial functions on $\bR^D$, $L$ the
polynomial vectorfields, and let
$S=g_{ij}x^i x^j/2$.
Then there is exactly one renormalized operation satisfying the
renormalization condition
$f\partial_i ._r g=f._r \partial_i g$, and there is exactly one solution
to the corresponding Frobenius-Schwinger-Dyson equation.
\end{thm}
\begin{pr} (We use formal power series $e^{px}$ for neater expressions).
Indeed, let us first prove that $e^{px}._r e^{qx}=K(p,q) e^{(p+q)x}$ for
some $K$:
$$\partial_i
(e^{px}._r e^{qx})
=
\partial_i._r
(e^{px}._r e^{qx})
=
(\partial_i._r e^{px})._r e^{qx} + e^{px}._r (\partial_i._r e^{qx})
$$
$$=
(\partial_i e^{px})._r e^{qx} + e^{px}._r (\partial_i e^{qx})
= (p+q)_i e^{px}._r e^{qx},$$
which implies the above form. Next note that
$(f._r g)\partial = f ._r (g\partial)$ which is true since both sides have the
same renormalized action:
$$[(f._r g)\partial]._r h = (f._r g)._r \partial h
=f._r (g._r \partial h)
=f._r (g\partial._r h)
=(f._r g\partial)._r h.$$
We may use this identity to show that $K(p,q)=e^{pq}$:
$$(\partial_q K(p,q))e^{(p+q)x}
=_{(Leibniz)}\partial_q (K(p,q) e^{(p+q)x})
-K x e^{(p+q)x}$$
$$=_{(Definition\;of\;K)}\partial_q (e^{px}._r e^{qx})-(e^{px}._r e^{qx})
\partial S$$
$$=_{(Previous\;identity)}e^{px}._r (e^{qx}\partial S) - (e^{px}._r e^{qx}
\partial) S$$
$$=_{(Compatibility\;with\;S)}
e^{qx}\partial ._r e^{px}=e^{qx}._r \partial e^{px}=pK(p,q)e^{(p+q)x}.$$
So that $K(p,q)=e^{pq}$.
This in turn determines $X._r f$ by using the renormalization condition, and
$f._r X$ by the identity $f._r (g\partial_i)=(f._r g)\partial_i$.

We will now show that indeed $L=Der(A,._r)$.
By taking derivatives,
$x^i._r x^j= g^{ij}+x^ix^j$, and $A$ is $._r$-generated by the $x^i$'s.
Therefore any derivation $D_r$ of $._r$ equals $D_r(x^i)._r\partial_i._r$,
which is indeed of the form $X._r$ with $X\in L$, since
$D_r(x^i)._r\partial_i$ is in $L$.
Finally, $[X,_r Y]$ is determined by $[X,_r Y]._r g=
X._r Y._r g-Y._r X._r g$.

As for the expectation values, they follow from the operator product,
but can also be computed directly:
$$\partial_{p_0^i} \langle e^{p_0 x}, .. ,e^{p_n x}\rangle_r
=\langle e^{p_0 x} \partial_{x^i} S, e^{p_1 x}, .. ,e^{p_n x}\rangle_r$$
$$=\sum_{j=1}^n \langle e^{p_1x}, .., e^{p_0 x} \partial_i._r e^{p_j x}, ..
e^{p_n x}\rangle_r$$
$$=\sum_{j=1}^n \langle e^{p_1x}, .., e^{p_0 x}._r \partial_i e^{p_j x}, ..
e^{p_n x}\rangle_r$$
$$=(p_1+...+p_n)_i\langle e^{p_0 x}, .. ,e^{p_n x}\rangle_r.$$
This determines $\langle.\rangle_r$
\end{pr}
\subsection{Example 2: Renormalization conditions for Schroer's Lagrangian.}
Schroer's Lagrangian \cite[Formulae 24,30]{schroer} is given by
$$\cL^\lambda(\psi,\bar \psi,\phi):=\cL_0(\phi)
+[\bar\psi(\gamma^\mu\partial_\mu -im)\psi
+i\lambda \bar\psi \gamma^\mu \psi\partial_\mu\phi]
dx^1..dx^D,$$
where $\cL_0(\phi)$ is some solved Lagrangian for $\phi$, say quadratic.
For ease of computation we consider such a Lagrangian to be an element
of the symmetric algebra on even symbols $\phi(x)$ and odd symbols
$\psi^A(x),\bar\psi^A(x)$, and their derivatives. $\gamma_\mu$ is
a symmetric Dirac matrix $\gamma_\mu^{AB}=\gamma_\mu^{BA}$.
Schoer's interest in this Lagrangian was raised by the fact that
it is nonrenormalizable by powercounting,
but exactly solvable by the identity
$\cL^\lambda(\psi,\bar \psi,\phi)=\cL^0
(e^{i\lambda\phi}\psi,e^{-i\lambda\phi}\bar \psi,\phi)$.
\subsubsection{Naive functional integral solution.}
In naive functional integral notation, the solution is easy: If $\cO$
is a function of $(\psi,\bar\psi,\phi)$, then
$$\langle \cO \rangle^\lambda
= \int D\phi D\psi D\bar\psi
e^{-S^\lambda(\psi,\bar\psi,\phi)}\cO(\psi,\bar\psi,\phi)
= \int D\phi D\psi D\bar\psi
e^{-S^0(e^{i\lambda\phi}\psi,e^{-i\lambda\phi}\bar\psi,\phi)}
\cO(\psi,\bar\psi,\phi)$$
$$= \int D\phi De^{-i\lambda\phi}\psi De^{i\lambda\phi}\bar\psi
e^{-S^0(\psi,\bar\psi,\phi)}\cO(e^{-i\lambda\phi}\psi,e^{i\lambda\phi}
\bar\psi,\phi)$$
$$= \int D\phi D\psi D\bar\psi
e^{-S^0(\psi,\bar\psi,\phi)}\cO(e^{-i\lambda\phi}\psi,e^{i\lambda\phi}
\bar\psi,\phi)
=\langle \cO(e^{-i\lambda\phi}.,e^{i\lambda\phi}.,.)\rangle^0,$$
which e.g. for the two-point function reads
$$\langle \psi^A(x)\bar\psi^B(y)\rangle^\lambda
=
\langle e^{-i\lambda\phi(x)}\psi^A(x)
e^{i\lambda\phi(y)}\bar\psi^B(y)\rangle^0,$$
which is however nonsense since it contains unregularized composite
operators. Schroer's expectation value is the following modification of this
expression in case $\cL_0$ is quadratic:
$$\langle \psi^A(x)\bar\psi^B(y)\rangle_{Schroer}^\lambda:
=
\langle N(e^{-i\lambda\phi(x)})\psi^A(x)
N(e^{i\lambda\phi(y)})\bar\psi^B(y)\rangle^0$$
$$=e^{\lambda^2 \langle \phi(x)\phi(y)\rangle}
\langle \psi^A(x)\bar\psi^B(y)\rangle^0,$$
where $N$ is normal ordering for the Lagrangian $\cL_0$. Or more explicitly
if $\cL_0(\phi):=(\partial_\mu\phi)^2/2$, with $k_D:=Vol(S^{D-1})^{-1}$:

\[\langle \psi^A(x)\bar\psi^B(y)\rangle_{Schroer}^\lambda=
\langle \psi^A(x)\bar\psi^B(y)\rangle^0
\times
\left\{
\begin{array}{ll}
\exp\{k_D \lambda^2 |x-y|^{2-D}\}  & D\neq 2 \\
|x-y|^{-k_D \lambda^2} & D=2
\end{array}
\right. \]
where we see an essential singularity in $|x-y|$ as soon as
$D>2$, which is in agreement with powercounting.
\footnote{
Powercounting \cite[Section V]{dyson.smat} is the order by order in $\lambda$
analysis of the divergence of the Fourier transform
of the expectation values. The result is that the divergence of the Fourier
transform becomes worse as the order of $\lambda$ increases if the
so-called dimension of the corresponding nonquadratic term in the Lagrangian
is bigger than zero;
The bad behaviour criterion for Schroer's Lagrangian reads
$0<[\bar\psi \gamma^\mu \psi\partial_\mu\phi dx^1..dx^D]
=2[\psi]+[\partial_x]+[\phi]+D[dx]$, where $[\partial_x]:=-[dx]:=1$, and
$[\psi],[\phi]$ are defined as {\it half} the power of
singularity of the unregularized Gaussian
{\it two}-point functions,
in this case $[\phi]={1\over 2}(D-2)$, $[\psi]={1\over 2} (D-1)$,
giving $0<(D-1)+1+(D/2-1)-D=D/2-1$.
We can see this directly from the exact solution, since the essential
singularity is present for $D>2$ only.
More generally powercounting asserts that regularized
Fouriertransformed expectation
values of a finite-dimensional
linear family of Lagrangians with a term of positive dimension
are not renormalizable within this family,
even if one includes all possible Lagrangians $\cL_i$ with $[\cL_i]\leq 0$
in this family (the so-called counterterms: They are designed to possibly
absorb divergences in their linear pre-factor, and are restricted such as to
not worsen the divergences that one wishes to absorb). Here it is understood
that the family should remain renormalizable when using
\underline{different} regulators
of the Gaussian two-point functions, like for normal ordering. E.g.
if the regulated Gaussian part was just
$\phi\Delta(1+(\Delta/\Lambda)^9)\phi$, then a simple counterterm would be
$\phi\Delta^9 \phi$, which after renormalization would amount to keeping
$\Lambda$ finite. However the answer would depend on $k$. Thus, in our
language it is better to think of the space of cutoffs to be very big
in order to include all $k$'s, and the space $\cB$ to be the linear family
of Lagrangians mentioned above.

The powercounting criterion
seems to be really about the renormalizability of the
Fourier transform \underline{only}; From Schroer's example we see that
it doesn't give us any information
on the existence of the functional integral in position space. This
asymmetry originates in the fact that a Lagrangian which is local in
$x$'s is non-local in $p$'s.
We may however use the momentum
powercounting tot draw conclusions on the \underline{type}
of $|x-y|$ singularities that
are to be expected, even if the theory is not exactly solvable. (Assuming that
the expectation values exist in $x$-space).}

\subsubsection{More sensible solution.}
The above formulation is unsatisfactory in that the problem was defined
during its own solution.
With the notions that we have introduced in the previous sections, it will
now not be very difficult to state the exact renormalization conditions
which single out Schroer's solution as a solution to the
Frobenius-Schwinger-Dyson equation:
\begin{thm}
There is exactly one solution of the Frobenius-Schwinger-Dyson
equation for Schroer's Lagrangian $\cL^\lambda$, satisfying the
renormalization condition $\forall_{f,g}\; fX._r g=f._r X g$, for
$X$ the following derivations:
$$e^{i\lambda \phi} {\delta\over \delta \psi^A};\;
e^{-i\lambda \phi} {\delta\over \delta \bar \psi^A};\;\;
{\delta\over \delta \phi}-i\lambda \{
\bar \psi^A
{\delta\over \delta \bar \psi^A}
-\psi^A
{\delta\over \delta \psi^A}\}.$$
\end{thm}
\begin{pr}
Indeed, we will prove in a moment that
under the map $M:\psi\mapsto e^{i\lambda\phi}\psi, \phi\mapsto \phi$
between symbolic algebras,
these vectorfields get mapped to
${\delta\over \delta \psi^A}$, etc.
Now we already know that there is a unique solution
for Gaussian actions of the Frobenius-Schwinger-Dyson equation such that
these last vectorfields commute with the $._r$ operation.
Thus, pulling everything back by the invertible $M$, we see that there is
a unique solution of the Schroer Lagrangian satisfying the pulled-back
renormalization conditions.
Finally let us prove the statement: We have
$$M({\delta\over \delta \psi^A})(\cO)
=M({\delta\over \delta \psi^A})MM^{-1}(\cO)
:=M({\delta\over \delta \psi^A}M^{-1}(\cO)).$$
Therefore,
$$M({\delta\over \delta \psi^A})(\phi)=
M({\delta\over \delta \psi^A}M^{-1}(\phi))=0,$$
$$M({\delta\over \delta \psi^A(x)})(\psi^B(y))=
M({\delta\over \delta \psi^A(x)}e^{-i\lambda \phi(y)}\psi^B(y))
=e^{-i\lambda \phi(y)}\delta(x-y) \delta_B^A.$$
I.e. $M(\delta/\delta \psi^A(x))=
e^{-i\lambda \phi(x)} \delta/\delta \psi^A(x)$,
leading to the first two formulae; as for the third:
$$M({\delta\over \delta \phi(x)})(\phi(y))=\delta(x-y),$$
$$M({\delta\over \delta \phi(x)})(\psi^A(y))
=M({\delta\over \delta \phi(x)}e^{-i\lambda \phi(y)}\psi^A(y))$$
$$=-i\lambda \delta(x-y) M(e^{-i\lambda \phi(y)} \psi^A(y))
=-i\lambda \delta(x-y) \psi^A(y).$$
So that $M(\delta/\delta \phi(x))
=
\delta/\delta \phi(x) -i\lambda \psi^A(x)
\delta/\delta \psi^A(x)
+i\lambda\bar\psi^A(x)
\delta/\delta \bar\psi^A(x).$
\end{pr}
\section{Acknowledgments.}
I would like to thank Prof.Dr.R.H.Dijkgraaf and Dr.H.Pijls for supervising
my work. Further thanks go to Dr.A.A.Balkema, Dr.E.van den Heuvel,
and Prof.Dr.J.Smit for discussions.
This work was financially supported by ``Samenwerkingsverband
Mathematische Fysica van de Stichting FOM en het SMC.''
\newpage
\appendix
\section{Functional integration.}
\label{fintsec}
\subsection{The general idea.}
\subsubsection{Example.}
This section is an introduction to the definition of integration over function
spaces, possibly infinite dimensional. One may think of such an integral
as a limit of finite dimensional integrals, like:
$$
\langle
x^7 x^7 x^5 x^5
\rangle
=\lim_{n\rightarrow \infty}
{\int e^{-(x^1 x^1 + .. + x^n x^n)/2} x^7 x^7 x^5 x^5 dx^1 ..dx^n
\over
\int e^{-(x^1 x^1 + .. + x^n x^n)/2} dx^1 ..dx^n}=1.$$
This limit exists since the expression is independent of $n$ if $n>7$.
Note however that the limit of the unnormalized integrals equals
$\lim_{n\rightarrow\infty} \sqrt{2\pi}^n=\infty$.

The above formula
can be seen as a normalized version of an integral over $\bR^\bN$. We
will also be concerned with cases of a ``continuum of integral signs'', i.e.
for example integrals over $\bR^\bR$. Trying to define them in a way
analogous to the above would need a discretization of that continuum, which
we will avoid by working with defining properties:

\subsubsection{Definition by properties rather than by construction.}
Before embarking on the definition of functional integration,
let us make a remark on usual
integration by noting that $\int_a^b f(x) dx$, in the early days, was defined
as $F(b)-F(a)$ where $F$ was a solution of $F^\prime=f$. Only much later
was the integral expression defined as the limit of Riemann sums, which at the
same time proved existence of an $F$ satisfying $F^\prime=f$,
even if one could not find it in terms of standard functions.
We will take the same approach to functional integration, stating it as a
problem, analogous to $F^\prime=f$, deferring the existence and
uniqueness questions of this problem to the indefinite future.

Note however that we cannot just take the defining property to be
the infinite dimensional limit of $F^\prime=f$, since
\begin{enumerate}
\item This would basically
define functional integration to be the \underline{unnormalized}
expression $\lim_{n\rightarrow \infty} \int dx^1 ..dx^n$, which we want
to avoid in the above example.
\item We will only be interested in
\underline{indefinite} integrals, so that we will have no use of the
values $F(x)$ other than $F(\infty)-F(-\infty)$. 
\item \underline{Gaussian} integrals (corresponding to the description of
the least complicated physical situations)
would in this way not be particularly
easy since the primitive of $e^{-x^2}$ is hard to find, and finally,
\item Since the integrals have to be normalized,
the interest is in the comparison of
\underline{different integrands}, say $\langle 1 \rangle$ and
$\langle x^2 x^2\rangle$, not the
integral of one single integrand.
\end{enumerate}
To that end we will
concentrate on the properties up to normalization of the
linear map $f\mapsto [f]$ below,
rather than on the function $F(a):=\int_0^a e^{-S} dx$.
\subsection{The finite dimensional Schwinger-Dyson equation.}
\subsubsection{The Schwinger-Dyson equation on $\bR^n$.}
For fixed
$S:\bR^n\rightarrow \bR$,
consider the following linear functional:
$$f\mapsto [f]:=\int_{\bR^n} f e^{-S} dx^1..dx^n,$$
where $f$ and $S$ are restricted such that it is well defined, and such that
upon partial integration boundary terms are zero. In that case the
functional satisfies three properties:
\begin{enumerate}
\item $f>0\Rightarrow [f]>0$. (Positivity).
\item $[1]=\int e^{-S} dx^1 .. dx^n$. (Normalization).
\item $\forall_{i,f}\;[\partial_i(S) f]=[\partial_i f]$.
(Schwinger-Dyson equation).
\end{enumerate}
Indeed, for the last point:
$$0=\int \frac{\partial}{\partial x^i}(e^{-S} f) dx^1...dx^n
=\int e^{-S} (-\partial_i(S)f +\partial_i f) dx^1...dx^n
=[\partial_i f]-[f\partial_i S].$$
The interest of this equation is that it can be generalized to infinite
dimensions, by substituting functional derivatives for the partial
derivatives \cite[formula 45]{feynman}.
We set $\langle f\rangle:=[f]/[1]$.
\subsubsection{Uniqueness argument.}
An important remark to be made on
the Schwinger-Dyson equation combined with positivity and
normalization is that in finite dimensions it allows for one solution
at most;
\begin{thm}
\footnote{
\label{measurefoot}
I thank Dr.A.A.Balkema and Dr.E.van den Heuvel
for providing me with the main ideas in this proof.
A faster but not very rigourous
argument goes as follows:
Since $\langle.\rangle$ is positive, one expects it to be given by a
positive weight $e^{-P}$:
$\langle f\rangle=\int f e^{-P}dx$. Since $\langle.\rangle$
now satisfies the Schwinger-Dyson equation for both $S$ and $P$, we have
$\forall_f \langle \partial_i(S-P) f\rangle=0$, which by positivity gives
$\partial_i(S)=\partial_i(P)$, or $P=S+c$, so that
$\langle f\rangle=K.\int f e^{-S}dx$,
so that $\langle .\rangle$
is determined up to a positive scalar, which is in turn
determined by the normalization condition.}
Let $S$ be a polynomial on $\bR$ such that $\int_{\bR} e^{-S(x)} dx$
converges.
Let
$C_{Pol}^\infty:=\{f\in C^\infty(\bR,\bR):
\exists_{Polynomial\;p} |f|<p\}$,
and let $\langle.\rangle:C_{Pol}^\infty\rightarrow \bR$ be a
normalized linear positive functional satisfying
$ \langle S^\prime f\rangle = \langle f^\prime \rangle$.
For $f\in C_{Pol}^\infty$, let $[f]:=\int fe^{-S} dx/\int e^{-S}dx$.
Then $\forall_{f\in C_{Pol}^\infty}\;\langle f \rangle=[f]$.
\end{thm}
\begin{pr}
\begin{enumerate}
\item First we prove that there is a $K$ such that
$\forall_{f\in C_{c}^\infty}\;
\langle f\rangle = K[f]$,
where
$C_c^\infty$ denotes the $C^\infty$-functions
with compact support.
By Riesz' representation theorem there is
a measure $d\mu$ on $\bR$ such that for all $f\in C_c^\infty$:
$\langle f\rangle = \int f d\mu$.

We claim that $\int_{\bR} d\mu$ and
$\forall_{t\in {\bR}} \tilde S(t):=\int_t^\infty S^\prime d\mu$ exist.
We will then show using the Schwinger-Dyson equation that $d\mu$
satisfies an identity involving $\tilde S$.
\begin{enumerate}
\item We have
$\int_\bR d\mu=\sup_{f\in C_c^\infty; f\leq 1} \int f d\mu
=\sup_{f\in C_c^\infty; f\leq 1} \langle f \rangle \leq
\langle 1 \rangle = 1,$ which proves the first existence.
\item
Next, if
$f\geq 0$ has compact support and $f_\lambda(x):=f(x-\lambda)$, then
by dominated convergence
$\lim_{\lambda\rightarrow \infty} \int f_\lambda d\mu =0$.
Using this, we now prove that
$\int_t^\infty S^{\prime} d\mu$ exists: Indeed we may assume
that $S^\prime>0$ on $[t-\epsilon,\infty)$, so that
$$\int_t^\infty
S^\prime d\mu=\sup_{0\leq f \leq 1;\;f\in C_c^\infty}
\int_t^\infty S^\prime f d\mu < sup_\lambda
\int_{t-\epsilon}^\infty S^\prime g_\lambda d\mu,$$
where $g_\lambda$ are more and more stretched bump functions as
$\lambda$ increases: $supp(g_\lambda)=[t-\epsilon,\lambda+\epsilon]$, and 
$g_\lambda([t,\lambda])=1$.
$$=\lim_{\lambda\rightarrow \infty}
\int_{t-\epsilon}^\infty S^\prime g_\lambda d\mu
=\lim_{\lambda\rightarrow \infty} \langle S^\prime g_\lambda \rangle
=\lim_{\lambda\rightarrow \infty} \langle g_\lambda^\prime \rangle
=\lim_{\lambda\rightarrow \infty} \int g_\lambda^\prime d\mu,$$
which exists.
\end{enumerate}
Let us now derive the identity for $d\mu$ which follows from the
Schwinger-Dyson equation: For $f\in C_c^\infty$:
$$\int_{-\infty}^\infty f^\prime d\mu=
 \langle f^\prime \rangle
=\langle S^\prime f\rangle
=\int_{-\infty}^\infty S^{\prime} f d\mu$$
$$
=\int_{x=-\infty}^\infty S^\prime (x)
\int_{t=-\infty}^x f^\prime (t) dt d\mu(x)
=\int_{-\infty}^\infty \int_{-\infty}^\infty
\phi(x,t) dt d\mu(x).$$
where $\phi(x,t):=S^\prime(x) f^\prime(t) (x\geq t)$.
We claim that we can interchange the integrals: By Fubini it suffices to
check that the repeated integral of $|\phi|$ exists.
Indeed we already know that $G(t):=\int_t^\infty |S^\prime|(x) d\mu(x)
=\int_{-\infty}^\infty |S^\prime (x)| (x\geq t) d\mu(x)$ exists.
Therefore $\int \int |\phi| d\mu(x) dt = \int |f^\prime| G dt<\infty$, since
$|f^\prime|$ has compact support. Therefore, interchanging the integrals:
$$=
\int_{t=-\infty}^\infty
[\int_{x=t}^\infty S^\prime (x) d\mu (x)]
f^\prime (t) dt =
\int_{t=-\infty}^\infty \tilde S(t) f^\prime(t) dt.$$
I.e. the functional $I(g):=\int g (d\mu-\tilde S dx):C_c^\infty\rightarrow
\bR$ satisfies $I(C_c^{\infty\prime})=0$, i.e. is a linear
function on $C_c^\infty/C_c^{\infty\prime}$, which is one-dimensional.
The dual of $C_c^\infty/C_c^{\infty\prime}$ is therefore also one-dimensional,
and it is generated by $g\mapsto \int g dx$. Therefore:
$$\exists_k \forall_{g\in C_c^\infty}
\int g d\mu = \int g(\tilde S+k) dx=:\int g d\nu.$$
This implies $d\mu=d\nu$ if $d\mu(K)$ and $d\nu(K)$ are finite for compact
$K$'s. Indeed if $f\in C_c^\infty$, $0\leq f \leq 1$, $f|_K=1$ then
$d\mu(K),d\nu(K)\leq \int_\bR fd\nu =\int_\bR f d\mu=\langle f \rangle \in\bR$.
Therefore $d\mu = (\tilde S+k) dx$, so that by the definition of $\tilde S$:
$$\tilde S(t)=\int_t^\infty S^\prime(x) (\tilde S(x)+k) dx,$$
so that $\tilde S$ is differentiable, and $-(\tilde S+k)^\prime=
S^\prime(\tilde S+k)$,
giving $\tilde S+k=\tilde k e^{-S}$, so that $d\mu=\tilde k e^{-S}dx$.
\item Next let us prove that $K=1$. To that end we have to fit the result
$\langle f\rangle=K[f]$
for $f$ of compact support with $\langle 1\rangle = [1]$.
Both $f\in C_c^\infty$ and $f=1$ can seen as functions on the circle
$f\in C^\infty(S^1)$. In order to compare the two, we
apply Riesz' theorem for functions on the circle, concluding that
there is a measure $d\rho$ on $S^1$ such that for $f\in C^\infty(S^1)$
we have $\langle f \rangle = \int_{S^1} f d\rho = d\rho(\infty) f(\infty)
+ \int_{-\infty}^\infty f d\rho$. Restricting this to functions with
compact support in $\bR$
we see that $d\rho|_\bR=d\mu=\tilde k e^{-S}dx$.
I.e. for $f$ on the circle $\langle f \rangle = d\rho(\infty) f(\infty)+ K[f]$.
Taking $f=1$ gives $1= d\rho(\infty) +K$, so that for functions on the
circle
$$\langle f \rangle = (1-K) f(\infty)+ K[f].$$
Therefore, if $f$ is such that $S^\prime f$ and $f^\prime$ extends to the
circle, then $0=(1-K)(S^\prime f-f^\prime)(\infty)$. To prove that $K=1$,
it suffices to find such a function such that $(S^\prime f-f^\prime)(\infty)
\neq 0$. Indeed take $f$ to be a $C^\infty$-function which equals
$1/ S^\prime$ in a neighborhood of $\infty$. Then
$f S^\prime\rightarrow 1$ and $f^\prime\rightarrow 0$,
giving $K=1$.
\item
\label{largepol}
Next define polynomials by $f_1:=1$ and $f_{n+1}=F_n.S^\prime$,
where $F_n^\prime = f_n$.
Then we have
$\langle f_n \rangle = [f_n]$:
Indeed this is clear for $f_1$, and by induction
we have
$$\langle f_{n+1}\rangle
=\langle F_n S^\prime \rangle
=\langle f_n \rangle
=[ f_n]=[ F_n S^\prime]
=[ f_{n+1}].$$
Since the $f_n$ get arbitrarily high degree as $n$ increases,
we see that for every $p\in C_{Pol}^\infty$, there is a polynomial $P>p$
such that $\langle P\rangle = [P]$.
\item We now prove the identity $\langle p \rangle=[p]$
for $p\in C_{Pol}^\infty$. By adding polynomials of point \ref{largepol}
to $p$, we may assume that $p>0$.
We will approximate $p$ by a function $p_\epsilon$
of compact support. Then we have
$$|\langle p \rangle -[p]|
\leq
|\langle p \rangle - \langle p_\epsilon \rangle|+
|\langle p_\epsilon \rangle -[p]|
=
|\langle p \rangle - \langle p_\epsilon \rangle|+
|[p_\epsilon] -[p]|.$$
I.e. it suffices to find a $p_\epsilon$ such that
$|\langle p \rangle - \langle p_\epsilon \rangle|$
and
$|[p_\epsilon] -[p]|$ are smaller than $\epsilon$.
Indeed by point \ref{largepol} pick a polynomial
$P>p$ such that $[P]=\langle P \rangle$. Since $[.]$ is given
by the exponential integral, there is a $P_\epsilon\in C_c^\infty$
such that $0\leq P_\epsilon \leq P$ and $[P-P_\epsilon]<\epsilon$. And
there is a $p_\epsilon\in C_c^\infty$ such that $0\leq p-p_\epsilon 
\leq P-P_\epsilon$
and $[p-p_\epsilon]<\epsilon$. Indeed this $p_\epsilon$ satisfies
$$|\langle p \rangle - \langle p_\epsilon \rangle|
=
\langle p - p_\epsilon \rangle
\leq
\langle P - P_\epsilon \rangle
=
[P - P_\epsilon]<\epsilon,$$
and
$$|[p_\epsilon] -[p]|
=[p-p_\epsilon]<\epsilon.$$
\end{enumerate}
\end{pr}
\subsubsection{Combinatorical case, normal ordering.}
In case the second derivatives of the action can be written as a polynomial
of the first derivatives, one may try to solve the Schwinger-Dyson
equation by combinatorics, by writing it as
$$\langle \partial_{i_1}(S)...\partial_{i_n}(S)\rangle =
\sum_{k=2}^n \langle \partial_{i_2}(S)..\partial_{i_1}\partial_{i_k}(S)..
\partial_{i_n}(S)\rangle .$$
This way of approaching the Schwinger-Dyson equation was studied
in somewhat greater generality in \cite{mirleau}.
The normal ordering operation is linked with such combinatorical solutions
as follows:
If $N$ is invertible, then
$\{I$ satisfies the Schwinger-Dyson equation $\Leftrightarrow
I(f)=ZN^{-1}(f)I(1)\}$,
where $Z$ denotes the projection of polynomial functions of the $S_i$'s
on their constant part, e.g. $Z(3+aS_1+b S_1 S_5)=3$.
\subsubsection{The Schwinger-Dyson equation for volume manifolds.}
\label{volmansec}
\begin{defn}
A volume manifold is a combination $(M,\mu)$, where $M$ is a manifold and
$\mu$ is a volume-form, i.e. a differential form of maximal degree
such that $\mu_m$ is nonzero in every point $m\in M$.
To any manifold is associated an associative algebra $A$, the real functions,
and a Lie algebra $L$, the vectorfields.
In addition to these objects, a volume manifold gives rise to a map
$\nabla:L\rightarrow A$ defined by
$\nabla(X)\mu:=L_X\mu.$
\end{defn}
Defining $[f]:=\int_M f \mu$, and assuming that $\partial M=0$,
we see that the identity $0=\int_M L_X(f\mu)$ can be phrased as
$$[X(f)]+[f\nabla(X)]=0,$$
which is the generalization of the Schwinger-Dyson equation for arbitrary
volume manifolds. Thus we see that the statement of the Schwinger-Dyson
equation does not need $\mu$, only $\nabla$. In fact:
\begin{thm} $\nabla$ above satisfies the following properties:
\label{vol_div_th}
\begin{enumerate}
\item It is closed: $\nabla([X,Y])=X(\nabla(Y))-Y(\nabla(X))$.
\item It is local: $\nabla(fX)=X(f)+f\nabla(X).$
\item $\nabla$ fixes $\mu$ up to multiplication by a locally constant function.
\end{enumerate}
\end{thm}
\begin{pr}
See \cite[Appendix E]{mirleau}.
\end{pr}
We might roughly state the above as follows: If we are not interested in
a particular normalization of the integral, then all the information
contained in $(M,\mu)$ is in $(A,\nabla)$.
Since furthermore infinite dimensional volume forms do not exist
it is useful to define an abstraction of a volumeform which still makes
sense in infinite dimensions:
\begin{defn}
By a formal volume manifold, we mean a combination $(A,\nabla)$, where
$A$ is a symmetric associative algebra with unit, and 
$\nabla$ is a divergence operator,
meaning that with $L:=Der(A)$, $\nabla:L\rightarrow A$, such that:
\begin{enumerate}
\item $\nabla([X,Y])=X(\nabla(Y))-Y(\nabla(X))$.
\item $\nabla(fX)=X(f)+f\nabla(X)$.
\end{enumerate}
\end{defn}
\subsection{Functional integration.}
Since finite dimensional integration amounts to finding positive  solutions
of the finite dimensional Schwinger-Dyson equation, we will define functional
integration as finding positive solutions of an infinite dimensional
Schwinger-Dyson equation. In this section we will be concerned with
stating the infinite dimensional Schwinger-Dyson equation, and will
make some remarks on the unicity of their solutions, and conjecture a
link with the theory of phase transitions.
\subsubsection{The Schwinger-Dyson equation for linear target spaces.}

In this section we will define what we mean by infinite dimensional
integration over linear function spaces. To that end we recall the
notion of directional differentiation in (possibly infinite
dimensional) linear spaces:
\begin{defn} Let $V$ be an $\bR$-vectorspace, let $v,w\in V$, and
let $f:V\rightarrow \bR$. If it exists, the derivative of $f$ at $w$ in
direction $v$ will be defined as
$$(\partial_vf)(w):=\partial_t|_{t=0} f(w+tv).$$
\end{defn}
In finite dimensions we have
$(\partial_v f)=v^i\partial_i f$, so that the Schwinger-Dyson equation can
be written as 
$\forall_{v\in\bR^n}\;\;[(\partial_v S)f]=[\partial_v f]$. This expression
can now be generlized to $v$ in infinite dimensional spaces, leading to the
following definition:
\begin{defn}
\label{fintdef}
Let $M$ be a manifold,
$V:=Map(M,\bR^n)$, and $\cO:=Map(V,\bR)$ (observables). Let $S\in\cO$, and
let $C$ be a normalization condition (like $[1]=1$), then by a functional
integral for $(S,C)$, we mean a linear map on (a subclass
\footnote{(It is not clear what type of conditions one might need
in the future here, but this doesn't
seem relevant at this stage of understanding).}
of) $\cO$,
satisfying:
\begin{enumerate}
\item $f>0\Rightarrow [f]>0$, (Positivity).
\item $\forall_{v:M\rightarrow \bR^n,\;compact\;support}\;\;
[(\partial_v S)f]=[\partial_v f]$. (Schwinger-Dyson equation).
\item $[.]$ satisfies normalization condition $C$. (Normalization).
\end{enumerate}
\end{defn}
The reason for which we restrict to compact support is that if $\partial M=
\emptyset$ then by partial integration this is equivalent to
$[(\delta S/\delta \phi(x))f]=[\delta f/\delta \phi(x)]$, which is the
usual formulation. Note that if $x\in \partial M$, this last equation will
not be true any more. Choosing for the first version makes it possible
to derive quantum mechanics from the Schwinger-Dyson equation: This involves
functional integrals with an interval $[0,t]$ as a base manifold, and
the derivation of the Schroedinger equation $(\partial_t)$ involves
manipulations at the boundary. Furthermore, we will see in a moment
that this restriction
to compact support
seems to be related to the existence of phases and phase transitions.
\subsubsection{Functional integration over nonlinear spaces.}

The generalization of the definition \ref{fintdef}
to the nonlinear case needs two
adaptations: First we have to specify the analogue of the vectorfields
that we used, second, we have to work with divergences $\nabla$ instead of
actions $S$.

As vectorfields for the Schwinger-Dyson equation, we propose to use
the vectorfields ${f\otimes Y}$, defined below, where $f$ is a function
with compact support on $M$ and $Y$ is a vectorfield on $N$; The flow of
this vectorfield in functionspace is defined as follows:
Let $\gamma:M\rightarrow N$, then
$(F_t^{f\otimes Y}(\gamma)):M\rightarrow N$ is defined by
$$(F_t^{f\otimes Y}(\gamma))(m):=F_{f(m)t}^Y(\gamma(m)).$$

It remains to specify an infinite dimensional divergence $\nabla$.
Given the usual identity $\nabla_S=\nabla_0-dS$ holding for the divergences of
the volume forms $\mu_S:=e^{-S}\mu_0$,
one might think that the only way to specify a divergence $\nabla_S$, given
that we are interested in some action $S$, is to specify $\nabla_0$
(chosen as naturally as possible), and to set $\nabla_S:=
\nabla_0-dS$. As explained
in the text, this is not the only possibility.
A more convenient way to specify $\nabla_S$
is by a regulator $N$ of integrands and vectorfields, and setting
$$\nabla_S(N(X)):=-N(X(S)).$$
In this way one performs two tasks at the same time: The specification of
the divergence $\nabla_S$, and the specification of UV regulators, which
is obligatory for nonlinear targetspaces anyway, since there is
no natural distinction between linear and composite integrands in that
case.
\subsection{Pure phases, phase regions, phase transitions.}
\subsubsection{An analogy between lattice theories and the
Schwinger-Dyson equation with compactly supported vector fields.}
In finite dimensions, the sketchy argument
of footnote \ref{measurefoot} leading to the idea that positivity,
normalization and Schwinger-Dyson equation have at most one solution
contained the implication $\forall_i\;\partial_i(S-P)=0
\Rightarrow S=P+c$. Since in
infinite dimensions we restrict to vectorfields with compact support,
an analogous implication would not be possible since we would have
$\forall_v \partial_v(S-P)=0\Rightarrow$ $S$ and $P$ differ by a function which
only depends on the germ \underline{at infinity}.

Furthermore, from the work of scientists intersted in the existence of phase
transitions in lattice models, see \cite[Section 4]{dobrushin},
\cite[section II.5]{froehlich}
we know that that different ``Gibbs states'' may exist for a certain fixed
Hamiltonian, these Gibbs states being determined by that Hamiltonian and
the conditions \underline{at the boundary}
of an ever-increasing finite subset of the
lattice. In that context, one may prove that the set of equilibrium
Gibbs states is convex and that each of its points is a convex combination
of the extremal points.
\footnote{An extremal point of a convex set $\cC$ is a point which cannot
be written as $ac_1+(1-a)c_2$ with $c_i\in\cC$ and $a\in (0,1)$.}
Pure phases are then defined as these extremal points,
the ones in between correspond to states in which these phases
coexist.

Having these considerations in mind, we may note that the set of normalized
positive solutions of the Schwinger-Dyson equation is a convex set,
i.e. if $I_1$ and $I_2$ are two such solutions then so is
$aI_1 + (1-a) I_2$ for $a\in [0,1]$, and
by analogy with the lattice definition of a pure phase, we are led to the
following:
\subsubsection{Definition of phase transitions in the context of the
Schwinger-Dyson equation.}
\begin{defn} Let $\cC_S$ be the set of positive normalized solutions of the
Schwinger-Dyson equation for the action $S$. $\cC_S$ is convex.
\begin{enumerate}
\item By a pure phase of $S$ we understand an extremal point
of $\cC_S$.
\item A one-parameter family $\lambda\mapsto S_\lambda$ of actions is said
to undergo a phase transition at $\lambda_0$ iff the number of pure phases
of $S_\lambda$ changes at $\lambda=\lambda_0$.
\item Two actions are said to be in the same phase region iff they can
be joined by a path which does not undergo a phase transition.
\end{enumerate}
\end{defn}
\begin{conj}
Still by analogy with lattice results: (For more precise formulations
see \cite[Section II.5]{froehlich})
\begin{enumerate}
\item Every element of $\cC_S$
can be written as a convex combination of extremal elements of $\cC_S$.
(I.e. extremal points are really \underline{part of} $\cC_S$).
\item For a large class of actions $S$,
 $\cC_{S/\hbar}$ has exactly one point if we take
$\hbar$ large enough.
\item For a large class of actions $S$,
 $\cC_S$ has exactly one point if the base manifold is compact.
\end{enumerate}
\end{conj}
\newpage
\end{document}